# Multimodal MRI-Ultrasound AI for Prostate Cancer Detection Outperforms Radiologist MRI Interpretation: A Multi-Center Study

Hassan Jahanandish[a,b], Shengtian Sang[c], Cynthia Xinran Li[d], Sulaiman Vesal[a,b], Indrani Bhattacharya[e], Jeong Hoon Lee[b], Richard Fan[a], Geoffrey A. Sonn[a,b], Mirabela Rusu[b,a]

[a]Stanford University, Department of Urology, 300 Pasteur Drive, Stanford, 94305, California, USA

[b]Stanford University, Department of Radiology, 300 Pasteur Drive, Stanford, 94305, California, USA

[c]University of Miami, Department of Computer Science, Coral Gables, 33124, Florida, USA

[d]Stanford University, Institute of Computational and Mathematical Engineering, 475 Via Ortega, Stanford, 94305, California, USA

[e]Dartmouth College, Department of Biomedical Data Science, 1 Medical Center Drive, Lebanon, 03756, New Hampshire, USA

## Abstract

Pre-biopsy magnetic resonance imaging (MRI) is increasingly used to target suspicious prostate lesions. This has led to artificial intelligence (AI) applications improving MRI-based detection of clinically significant prostate cancer (CsPCa). However, MRI-detected lesions must still be mapped to transrectal ultrasound (TRUS) images during biopsy, which results in missing CsPCa. This study systematically evaluates a multimodal AI framework integrating MRI and TRUS image sequences to enhance CsPCa identification. The study included 3110 patients from three cohorts across two institutions who underwent prostate biopsy. The proposed framework, based on the 3D UNet architecture, was evaluated on 1700 test cases, comparing performance to unimodal AI models that use either MRI or TRUS alone. Additionally, the proposed model was compared to radiologists in a cohort of 110 patients. The multimodal AI approach achieved superior sensitivity (80%) and Lesion Dice (42%) compared to unimodal MRI (73%, 30%) and TRUS models (49%, 27%). Compared to radiologists, the multimodal model showed higher specificity (88% vs. 78%) and Lesion Dice (38% vs. 33%), with equivalent sensitivity (79%). Our findings demonstrate multimodal AI's potential to improve CsPCa lesion targeting during biopsy and treatment planning, surpassing current unimodal models and radiologists; ultimately improving outcomes for prostate cancer patients.

## 1. Introduction

Prostate cancer is the second-most prevalent cancer in men worldwide[1] and remains the cancer with second-highest mortality among American men[2]. The 5-year survival rate of prostate cancer patients has been reported to increase to 99% if the cancer is diagnosed in the early stages and treated while still localized or regional, in contrast with a reported 5-year survival rate of 30% for distant prostate cancer, where the cancer has spread to other organs and regions in the body[2]. Therefore, the early diagnosis of prostate cancer is of crucial importance.

Over the recent years, various initiatives have been helpful in enhancing the diagnosis and treatment of prostate cancer patients, including the use of pre-biopsy Magnetic Resonance Imaging (MRI) to target suspicious lesions during the transrectal ultrasound (TRUS)-guided biopsy procedure[3]. TRUS-guided biopsy relies on systematic sampling of the prostate at 12-14 locations which results in a 48% cancer detection sensitivity[3]. The systematic biopsy is improved by targeting suspicious lesions along with the systematic sampling locations during biopsy. Expert radiologists use pre-biopsy MRI scans to identify these suspicious lesions for targeting, which results in an increased overall diagnostic sensitivity of up to 88%[3]. The adoption of pre-biopsy MRI has seen a dramatic increase in recent years due to its superior diagnostic sensitivity, where its utilization increased significantly from 0.5% in 2007 to 35.5% in 2022[4]. However, the reliance on expert radiologist readings for the interpretation of pre-biopsy pelvic MR images remains a diagnostic bottleneck causing long wait times for patients, especially in rural areas with limited availability of expert radiologists[4,5]. Additionally, the interpretation of prostate MR images has been reported to substantially vary across radiologists, resulting in significant differences in cancer diagnostic yield[6].

Given the increasing utilization of pre-biopsy MRI, the inherent quality, and excellent soft tissue contrast of MR images, artificial intelligence (AI) methods that automate or assist with the detection, localization, or grading of prostate cancer on MR images have been extensively explored[7–9]. Various AI approaches have been proposed that use one or more MRI sequences, including T2-weighted (T2w) images, Apparent Diffusion Coefficient (ADC) images, Diffusion Weighted Images (DWI), and Dynamic Contrast Enhanced (DCE) images, to detect the presence of cancer, localize the cancerous lesions, and assess cancer aggressiveness[7–9]. Some of the recent work has explored radiomic-based machine learning (ML) approaches that rely on handcrafted features that might relate to cancer[10,11]. Most recent approaches utilize deep learning (DL) to train AI models that learn to directly identify cancer using MR image sequences as input[12–17]. These methods have shown great promise in achieving highly sensitive cancer detection and pixel-level localization performance. Specifically, a recent study reporting the outcome of the PICAI prostate cancer detection challenge reported that an AI system trained on 9129 patients was on average superior to a pool of 62 radiologists in detecting **C**linically **s**ignificant **P**rostate **Ca**ncer (CsPCa)[18].

While AI methods hold great promise in detecting and localizing cancer on MRI, the predicted lesions must still be accurately projected onto TRUS images for targeting during biopsy. However, TRUS-guided biopsies and pre-biopsy MRI scans are performed at separate times with a considerably different patient positioning, resulting in significant variations in the imaging field of view and the orientation of the acquired sequences[19,20]. Additionally, the shape of the prostate in TRUS images is altered due to deformation caused by the rectal insertion of the TRUS probe[19]. These factors make the accurate projection of the MRI-identified lesions to TRUS images a non-trivial challenge, typically addressed through cognitive registration by expert urologists or semi-automated registration facilitated by targeted biopsy systems[20,21]. However, such approaches are susceptible to registration errors, which have been reported to result in missing up to 20% CsPCa lesions during biopsy targeting[22–24]. On the other hand, the use of TRUS images as an independent input modality for AI-based prostate cancer detection and localization has been recently explored with promising preliminary results[25–32]. However, given the inherent differences between MR and ultrasound imaging modalities, the question remains whether combining both modalities can enhance the performance of AI-based cancer detection methods. While a recent study explored this concept through image-level classification[33], its findings offer limited utility for biopsy targeting, leaving this question largely unanswered.

In this study, we systematically assess the feasibility and impact of integrating MRI and TRUS images as inputs to a multimodal AI model compared to models that only use either of the modalities. We evaluate the model performances in three independent cohorts of patients from two institutions, totaling 3110 patients. We *hypothesize* that MRI and TRUS image sequences contain complimentary information, enabling a multimodal AI model to achieve higher performance in detecting and localizing CsPCa compared to models that solely rely on either of the two modalities. Additionally, we evaluate the performance of the AI models in detecting and localizing CsPCa in comparison with radiologists reading MR images during routine clinical care in an independent test cohort of patients. If confirmed, our study will provide the first evidence for the use of a large-scale multimodal AI model that integrates MRI and TRUS image sequences to enhance prostate cancer diagnostic accuracy.

## 2. Results

Our multimodal AI model is based on a 3D UNet backbone and integrates MRI (T2w, ADC, and DWI sequences) and TRUS image sequences as inputs to simultaneously segment the prostate gland, indolent cancer lesions, and CsPCa lesions. The model was evaluated using three independent test cohorts including two biopsy cohorts ($\boldsymbol{C}_{BX}^{Stanford}$ and $\boldsymbol{C}_{BX}^{UCLA}$) and one radical prostatectomy cohort ($\boldsymbol{C}_{RP}^{Stanford}$). These cohorts included a total of 1700 test studies from both internal and external institutions. Cancer ground truth labels were determined through pathology-confirmed targeted biopsy lesions or whole-mount histopathology following radical prostatectomy. We evaluated the diagnostic performance of the multimodal model against unimodal MRI and TRUS models, as well as radiologists, using standardized performance metrics (e.g., sensitivity, specificity, and Dice coefficient). Figure 1 shows an overview of the patient cohorts (Figure 1A) and the AI model architecture (Figure 1B). Detailed descriptions of the cohorts, imaging data, and AI model architecture are provided in the Methods section.

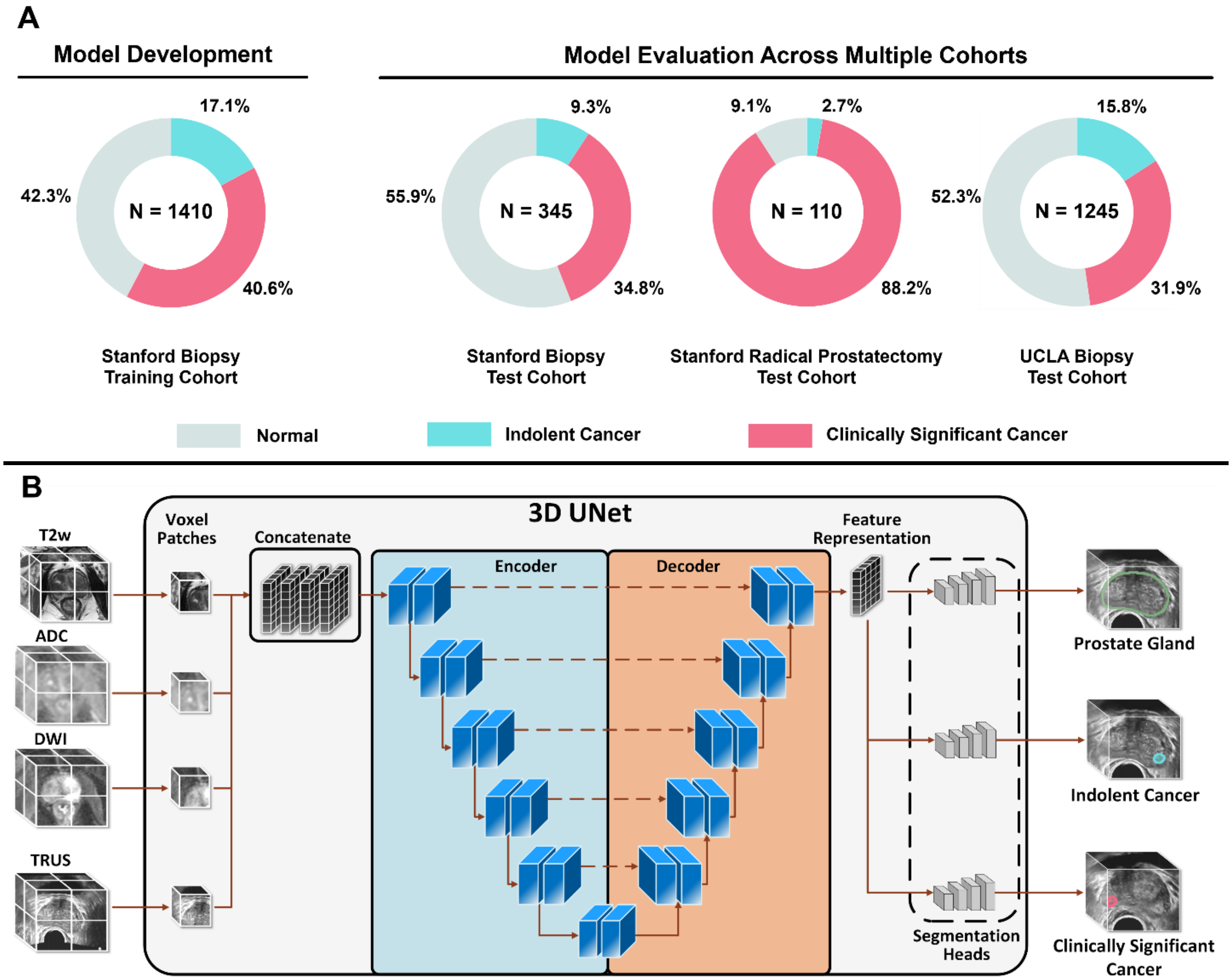

Figure 1: (A) Overview of the internal patient cohort used for model development and the three independent internal and external cohorts used for model performance evaluation. The Stanford Radical Prostatectomy test cohort was additionally used to compare AI cancer detection performance with radiologists. (B) illustrates the architecture of our multimodal deep learning AI model. The model uses all image sequences as input and independently segments the prostate gland, indolent cancer lesions, and clinically significant cancer lesions. The same network architecture was utilized for unimodal MRI and TRUS AI models except that voxel patches were extracted and concatenated from either the MRI or TRUS image sequences. The blue blocks indicate 3D convolutional layers followed by normalization and activation layers. Upward, downward, and dashed arrows show upsampling, downsampling, and skip connections.

**Multimodal Model vs. Unimodal Models:** Our proposed multimodal AI model was able to successfully learn to identify all three class labels and outperform both unimodal TRUS and MRI models in identifying CsPCa across all 1700 test cases. Table 1 details the performance of each model in each test cohort, where the multimodal model achieves a higher performance compared to both unimodal models across all test cohorts in all metrics except for specificity. Specifically, the multimodal model achieves significantly higher average sensitivity and overall Dice of 0.80 and 0.34 across all test cohorts compared to 0.73 and 0.23 for the MR model and 0.49 and 0.14 for the TRUS model. The multimodal model is less specific than the MR and TRUS models with a margin of 0.05 and 0.02, respectively, resulting in up to five extra false positive lesions compared to the MRI model for every 100 patients. Yet, the 0.97 negative predictive value (NPV) achieved by the multimodal model suggests the potential clinical utility of the model as a screening tool to avoid unnecessary biopsies as well as its highly sensitive performance in identifying cancer lesions.

Table 1: Multimodal AI model performance compared to unimodal models in identifying and localizing CsPCa

| | **Cohort** | **ROC** | **PR** | **Sensitivity** | **Specificity** | **NPV** | **Overall Dice** | **Lesion Dice** |
|---|---|---|---|---|---|---|---|---|
| **Unimodal TRUS Model** | $\boldsymbol{C}_{BX}^{Stanford}$ | 0.80 | 0.69 | 0.55 | 0.88 | 0.96 | 0.17 | 0.29 |
| | $\boldsymbol{C}_{RP}^{Stanford}$ | 0.73 | 0.63 | 0.48 | 0.90 | 0.85 | 0.14 | 0.26 |
| | $\boldsymbol{C}_{BX}^{UCLA}$ | 0.68 | 0.59 | 0.45 | 0.88 | 0.95 | 0.11 | 0.23 |
| | **Average:** | 0.74 | 0.64 | 0.49 | 0.89 | 0.92 | 0.14 | 0.27 |
| **Unimodal MRI Model** | $\boldsymbol{C}_{BX}^{Stanford}$ | 0.92 | 0.88 | 0.78 | 0.92 | 0.98 | 0.27 | 0.33 |
| | $\boldsymbol{C}_{RP}^{Stanford}$ | 0.90 | 0.83 | 0.72 | 0.93 | 0.93 | 0.22 | 0.30 |
| | $\boldsymbol{C}_{BX}^{UCLA}$ | 0.87 | 0.81 | 0.70 | 0.91 | 0.97 | 0.21 | 0.28 |
| | **Average:** | 0.90 | 0.84 | 0.73 | **0.93** | 0.97 | 0.24 | 0.30 |
| **Multimodal Model** | $\boldsymbol{C}_{BX}^{Stanford}$ | 0.93 | 0.89 | 0.81 | 0.85 | 0.98 | 0.37 | 0.45 |
| | $\boldsymbol{C}_{RP}^{Stanford}$ | 0.90 | 0.85 | 0.79 | 0.88 | 0.94 | 0.32 | 0.38 |
| | $\boldsymbol{C}_{BX}^{UCLA}$ | 0.90 | 0.85 | 0.79 | 0.87 | 0.98 | 0.34 | 0.42 |
| | **Average:** | **0.91** | **0.86** | **0.80** | 0.87 | **0.96** | **0.34** | **0.42** |

**Multimodal Model vs. Radiologists:** The performance of the multimodal model was compared to radiologist readings and evaluated against whole-mount pathology labels in $\boldsymbol{C}_{RP}^{Stanford}$ cohort. As detailed in Table 2, the multimodal model achieved the same sensitivity of 0.79 compared to radiologist readings while significantly outperforming them in other metrics including area under the receiver operating characteristic (ROC) and precision-recall (PR) curves, specificity, and Overall Dice by a margin of 0.11, 0.07, 0.10, and 0.05, respectively. When considering the model performance across all metrics, the unimodal MRI model also outperformed the radiologist readings, even though the radiologist readings had a higher sensitivity by a 0.07 margin. These highlight that a multimodal model has the potential to achieve a remarkably high cancer detection sensitivity comparable to expert world-class radiologists while predicting fewer false positives than radiologists.

**Qualitative Evaluation:** Qualitative review of the models' performance across various cases and cohorts followed a similar overall theme as the quantitative results, where the multimodal model outperformed the unimodal models in detecting CsPCa lesions as well as localizing them in majority of cases. The multimodal and unimodal MRI models achieved a comparable qualitative performance in detecting lesions when compared to radiologists. However, compared to radiologists, the multimodal model had fewer false positives, and it was particularly better at localizing CsPCa lesions. Figure 2 highlights three representative cases from $\boldsymbol{C}_{RP}^{Stanford}$ cohort, where the multimodal model successfully detected all the lesions and localized them with acceptable overlap. While the unimodal MRI model and radiologist readings detected the lesions in two of the cases, they still struggled with localizing their extent which might negatively affect the diagnostic accuracy if used for lesion targeting during TRUS-guided biopsy. Radiologist readings further missed the smaller lesion present in Case 1. Interestingly, while the unimodal MRI model and radiologist readings both failed to detect the CsPCa lesion present in Case 1, the unimodal TRUS model and the multimodal model both detected and successfully localized the lesion. This suggests the TRUS image sequence might have contained information not present in MR image sequences, which was successfully extracted by the multimodal model to enable the detection and localization of the lesion. Figure 3 further shows two representative cases from the external test cohort $\boldsymbol{C}_{BX}^{UCLA}$, where the performance of the multimodal model against the MRI model is illustrated in 3D across the prostate gland. While the multimodal model successfully detected the lesions in both cases, the MRI model failed to detect the small lesion present in Case 5. Both models showed over-localization and under-localization with respect to parts of the larger lesion present in Case 4. However, the multimodal model better captured the extent of the lesion, enhancing the chance of a successful targeted biopsy based on its localization compared to the MRI model.

Table 2: Multimodal AI model performance compared to radiologists in identifying and localizing CsPCa

| | **Model** | **ROC** | **PR** | **Sensitivity** | **Specificity** | **NPV** | **Overall Dice** | **Lesion Dice** |
|---|---|---|---|---|---|---|---|---|
| $\boldsymbol{C}_{RP}^{Stanford}$ **Cohort** | Unimodal TRUS Model | 0.73 | 0.63 | 0.48 | 0.90 | 0.85 | 0.14 | 0.26 |
| | Unimodal MRI Model | **0.90** | 0.83 | 0.72 | **0.93** | 0.93 | 0.22 | 0.30 |
| | Multimodal Model | **0.90** | **0.85** | **0.79** | 0.88 | **0.94** | **0.32** | **0.38** |
| | Radiologists | 0.79 | 0.78 | **0.79** | 0.78 | 0.93 | 0.27 | 0.33 |
| | **Improvement by AI:** | **0.11** | **0.07** | **0.0** | **0.10** | **0.01** | **0.05** | **0.05** |

**Failure Analysis:** The test cohorts included a total of 660 CsPCa, with ~15.3% of those being missed by the multimodal model. We found significant difference ($p < 0.0001$, two-way ANOVA test with Tukey's multiple comparisons test) in the volume of the missed lesions compared to the true positive lesions across all cohorts by any of the models or radiologist readings, as shown in Figure 4A. The median volume of the missed lesions by the multimodal model was 531 $mm^3$ (median diameter: 10.05 mm, and volume 90% confidence interval (CI): 254 $mm^3$) compared to a median volume of 1037 $mm^3$ (median diameter: 12.56 mm, and volume 90% CI: 189 $mm^3$) for the correctly predicted lesions across all cohorts. Figure 4B and 4C further show the distribution of true positive and false negative lesions based on their aggressiveness (GGs). Of all the lesions missed by the multimodal model across all cohorts, only 20% have a GG > 2, and only 7% have a

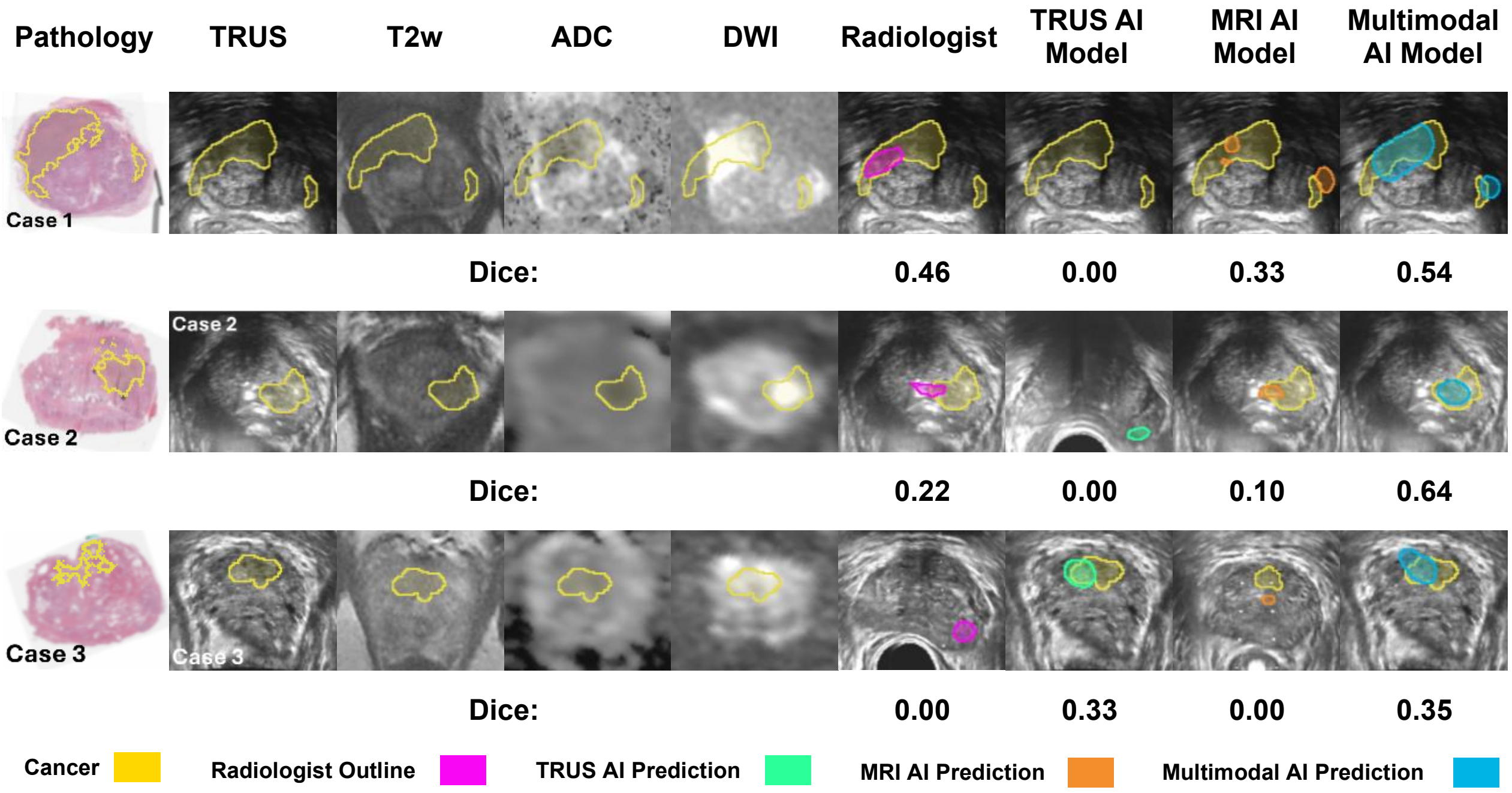

Figure 2: Qualitative performance of each AI model and radiologist readings against CsPCa ground truth (shown in yellow outlines) in three other representative cases from the independent test cohort $\boldsymbol{C}_{RP}^{Stanford}$. The TRUS and MRI image sequences that were used as input to various AI models are shown. While not explicitly used as input or ground truth for model training, the pathology images of the corresponding prostate slides are shown as the original reference for cancer ground truth. Note that pathology images have not been registered to either MRI or TRUS image sequences. In cases where a given AI model or radiologist reading did not find the cancer lesion, false positive lesions from other parts of the prostate gland are shown instead.

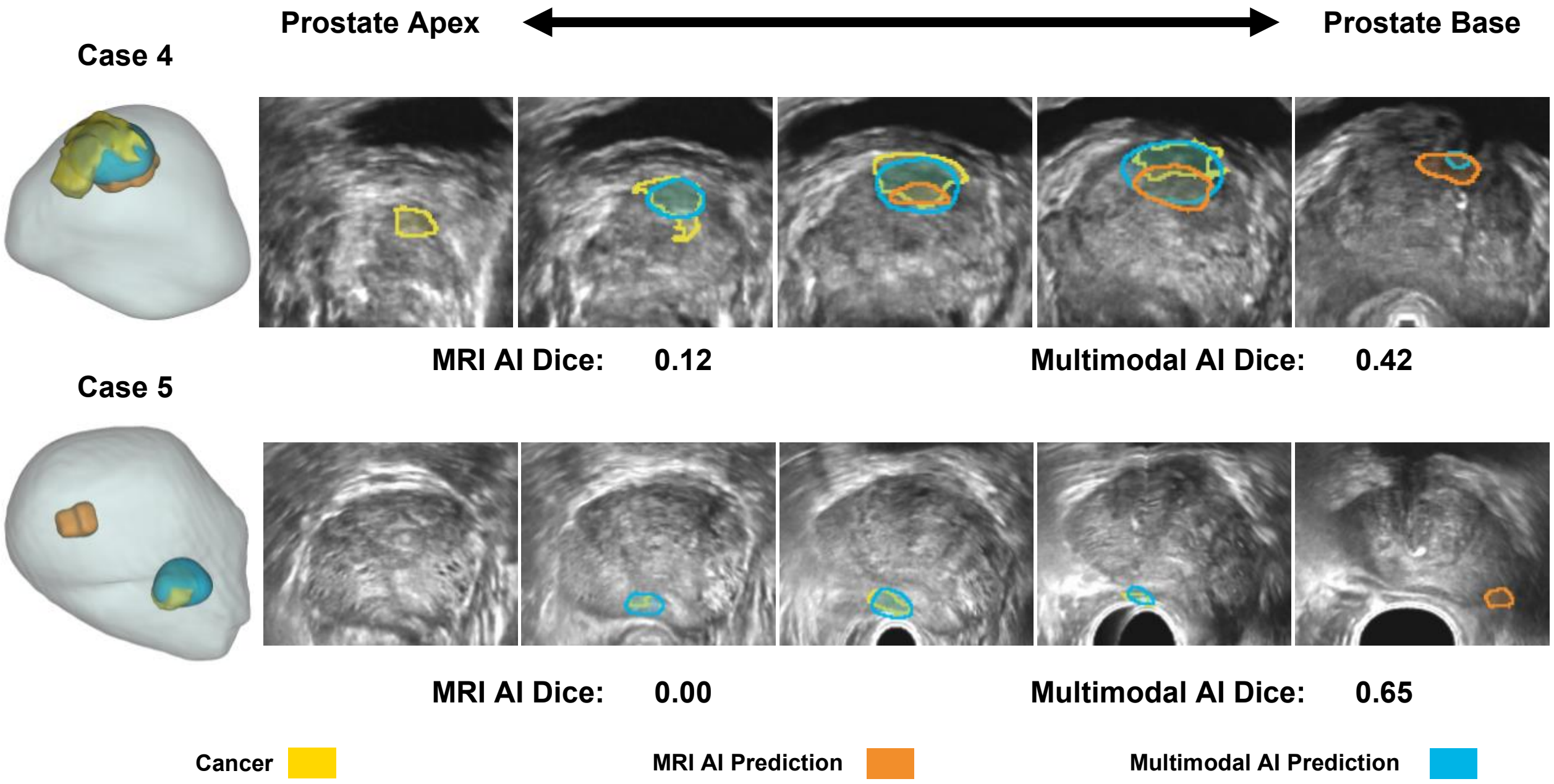

Figure 3: Qualitative performance of the Multimodal and MRI AI models shown in 3D against CsPCa ground truths (shown in yellow outlines) in two representative cases from the external test cohort $\boldsymbol{C}_{BX}^{UCLA}$. Various TRUS image slides have been shown across the sagittal plane through the prostate gland to highlight the AI model predictions across the entire extent of the lesions.

GG > 3. The models exhibit a similar behavior in the $\boldsymbol{C}_{RP}^{Stanford}$ cohort which is consistent with radiologist readings and no statistically significant difference was observed between the model predictions and radiologist readings ($p = 0.76$, two-way ANOVA test with Sidak's multiple comparisons test). However, the median volume of missed lesions in this cohort for radiologist readings was 1007 $mm^3$ (median diameter: 12.44 mm, and volume 90% CI: 410 $mm^3$) compared to the multimodal model at 795 $mm^3$ (median diameter: 11.49 mm, and volume 90% CI: 351 $mm^3$) (Figure 4A, right).

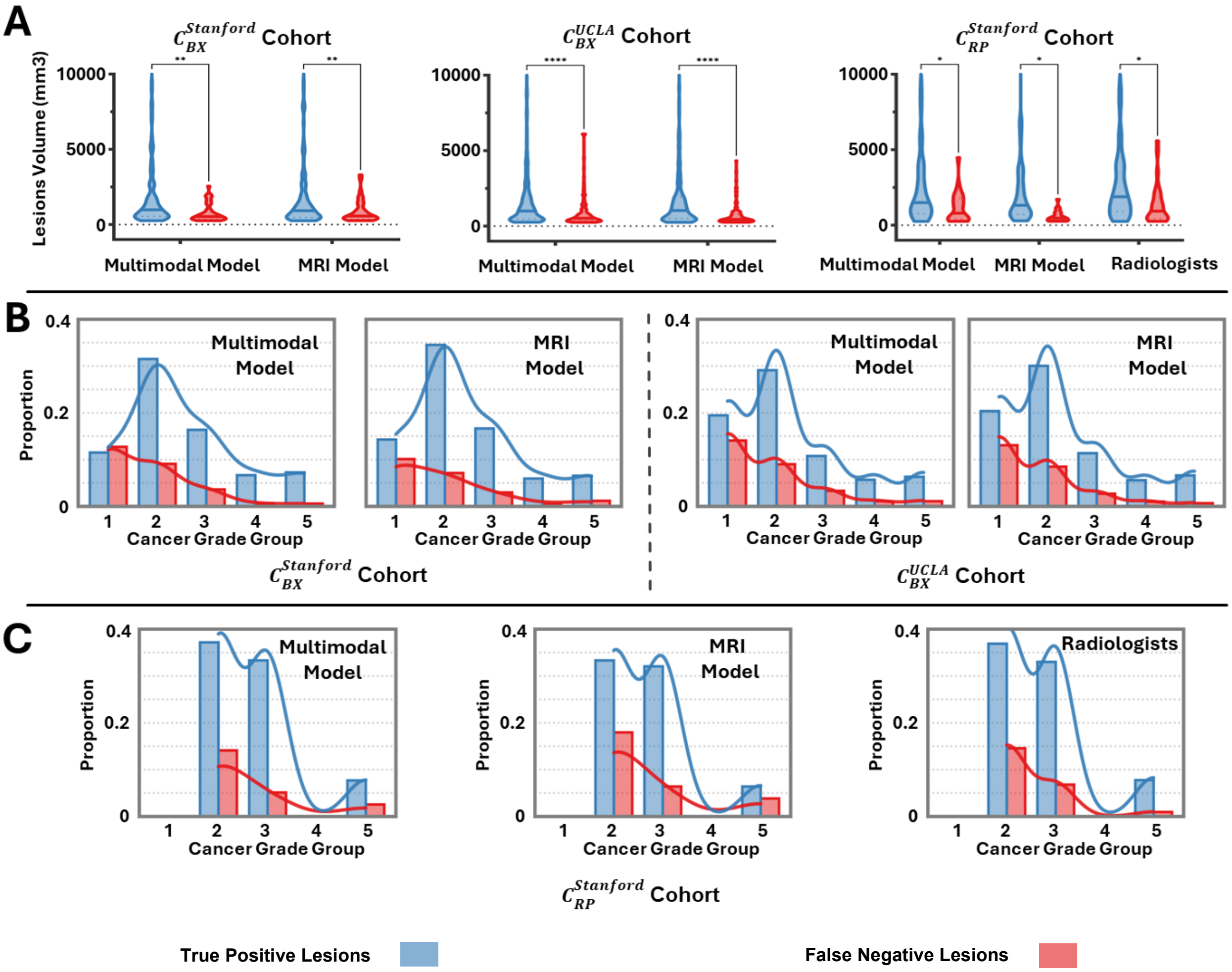

Figure 4: Analysis of the true positive and false negative lesions predicted by the AI models across various cohorts. (A) shows the consistent and statistically significant difference in the volume of the true positive and false negative lesions identified by the AI models as well as radiologists. (B) & (C) highlight the histogram distribution of the lesions in each group based on the cancer Grade Group (GG). $\boldsymbol{C}_{BX}^{Stanford}$ and $\boldsymbol{C}_{BX}^{UCLA}$ cohorts are shown in (B) and $\boldsymbol{C}_{RP}^{Stanford}$ cohort is shown in (C). *, **, and **** indicate $p < 0.05$, $p < 0.01$, and $p < 0.05$ for the significant difference between the true positive and false negative groups after multiple comparisons adjustment.

## 3. Discussion

In the present study, we demonstrated the feasibility of developing a large-scale multimodal AI model that utilizes both MR and TRUS image sequences as input modalities to identify and localize CsPCa directly on TRUS images for the first time. We further confirmed the hypothesis that MR and TRUS image sequences contain complimentary information that enables a multimodal AI model to outperform models that solely rely on either of the modalities by evaluating each model in three independent patient cohorts from two institutions. Additionally, we evaluated the performance of the multimodal AI model in an independent patient cohort compared to radiologists reading MR images during multidisciplinary routine clinical care and showed that its performance exceeds that of radiologists.

We evaluated the proposed multimodal AI model in three independent patient cohorts from two institutions, for a total of 1700 patients in the test cohorts. As detailed in Table 1, the multimodal AI model outperformed the other unimodal models in each test cohort. Specifically, the multimodal model achieved both statistically and clinically significant improvements in sensitivity as well as the Overall Dice and Lesion Dice compared to the unimodal MRI model in all three test cohorts. Those improvements ranged from 3% to 9% increase in sensitivity and 8% to 14% increase in Dice across the three test cohorts. Noteworthy, the multimodal model achieves the highest improvement in sensitivity compared to the unimodal MRI model in the external test cohort, $\boldsymbol{C}_{BX}^{UCLA}$, which results in identifying 40 additional CsPCa lesions in 33 patients across the cohort compared to the MRI model. While the multimodal model achieved a lower specificity compared to both unimodal models, it still achieved a great NPV value which was still marginally higher than the unimodal models, indicating the screening potential of the multimodal model.

We leveraged an independent cohort of radical prostatectomy patients, $\boldsymbol{C}_{RP}^{Stanford}$, with pixel-level whole-mount pathology ground truth to compare the performance of the proposed multimodal AI model with practicing radiologists reading pelvic MRI exams during multidisciplinary routine clinical care. As shown in Table 2, the multimodal model outperforms the radiologists by meaningful margins in most of our evaluation metrics. Specifically, the higher specificity of the multimodal model results in 41.5% fewer false positives at sextant level compared to radiologists. Interestingly, the unimodal MR model also achieves a higher overall rank than radiologists, which is consistent with the reported outcomes in Saha et al[18]. Furthermore, the multimodal model's performance in identifying more aggressive lesions was consistent with the radiologist readings (Figure 4C), which is promising towards a reliable Al-based prostate cancer diagnostic and screening tool.

Failure analysis of the multimodal AI model revealed that the missed lesions were significantly smaller in volume (median: 531 $mm^3$) compared to the true positive lesions (median: 1037 $mm^3$) across all cohorts. While less pronounced, radiologist readings exhibit a similar behavior to the multimodal AI models (Figure 4A). Noteworthy, recent clinical guidelines do not consider prostate cancer lesions smaller than 500 $mm^3$ to be clinically significant[18,34]. This suggests 44% of all the missed lesions by the multimodal AI model with GG $\geq 2$ would not be considered clinically significant according to this guideline. As further shown in Figure 4B, both the unimodal MRI and multimodal models have a similar lesion GG distribution across the cohorts, where almost all the missed lesions are concentrated in lower GGs indicating the reliability of the model in identifying more aggressive cancer lesions.

The strong performance achieved by the proposed multimodal AI model in identifying and localizing CsPCa as well as its success in outperforming both the unimodal MRI model and radiologist readings have significant long term clinical implications. While TRUS-guided systematic prostate biopsy remains the standard of care, pre-biopsy MRI followed by MR-targeted fusion TRUS biopsy is rapidly gaining adoption as the gold standard in the diagnostic pathway[4]. Yet, we demonstrated the feasibility of utilizing both MR and TRUS image sequences to develop a multimodal AI model trained on a large cohort of patients that outperforms the radiologist readings that dictate the diagnostic performance of the same gold standard. This suggests the potential of a multimodal AI approach to streamline prostate cancer diagnosis by removing the reliance on MRI-TRUS fusion biopsy devices and minimizing the need for expert radiologist interventions while maintaining the diagnostic performance.

Our study has a few limitations that should be addressed in future work. First, the AI models were evaluated on cohorts of retrospective data from two institutions. While the test cohorts include MRI data from multiple vendors and protocols, the TRUS image sequences shared the same vendor and probe. Though ultrasound imaging has inherently high inter-operator variability, future research should include TRUS data from multiple vendors, probe types, and frequencies to enhance the generalization potential of the models. Second, the labels used for training the models were pathology-confirmed radiologist outlines that were projected to TRUS images from MRI using the fusion system used for targeted biopsy. Therefore, these labels may bring registration errors and miss the lesions that radiologists may have missed or have been MRI invisible. To alleviate this issue, we included the independent test cohort with patients undergoing radical prostatectomy and used it to evaluate the model performance against pixel-level whole-mount pathology ground truth. Yet, these labels are only available in patients undergoing surgery in a subset of our cohorts, limiting their utility for model training. Future studies need to evaluate their approaches against these labels to better understand shortcomings of current technologies. Third, we used the established UNet architecture as proof of concept for the multimodal approach in this study. We anticipate that the use of more advanced AI model architectures and techniques will be effective in achieving better cancer detection and localization performance. Lastly, the performance of the multimodal model should be rigorously evaluated in prospective settings, both as an assistive tool for expert physicians and as an autonomous system.

To conclude, the present study assessed the feasibility of a novel approach in utilizing both MR and TRUS image sequences as input to a multimodal AI model for prostate cancer detection and localization. We demonstrated the effectiveness of this approach, confirming the hypothesis that MR and TRUS images capture complimentary information that could be integrated for enhanced performance. The proposed multimodal approach was evaluated with data from two institutions, where it outperformed unimodal models. Further, the multimodal AI model outperformed radiologists reading MR images during routine clinical care in an independent test cohort. The strong performance achieved by the proposed multimodal approach indicates its promise to significantly improve the diagnostic accuracy of clinically significant prostate cancer during biopsy, eventually enhancing patient outcomes.

## 4. Materials and Methods

*4.1. Description of the Data and Cohorts:*

**Cohorts:** The present study was approved by the Institutional Review Board (IRB) at Stanford University and included retrospective data from three patient cohorts consisting of a total of 3110 studies where each study represents a unique patient. The three patient cohorts were from two institutions, 1) $\boldsymbol{C}_{BX}^{Stanford}$ is an internal cohort consisting of 1755 patients who underwent MRI-US fusion targeted biopsy at Stanford Hospital, 2) $\boldsymbol{C}_{BX}^{UCLA}$ is an external cohort of publicly available data through The Cancer Imaging Archive (TCIA) from 1245 patients who underwent the same MRI-US fusion targeted biopsy at UCLA Hospital, 3) $\boldsymbol{C}_{RP}^{Stanford}$ is an internal cohort including 110 patients who underwent radical prostatectomy (RP) surgery at Stanford Hospital following their diagnosis of prostate cancer. The details of each cohort are highlighted in Figure 1A. For each study, three MR imaging sequences were captured during routine clinical care, including 1) T2w images, 2) ADC images, and 3) DWI images. Each study further included the TRUS image sequence that was captured using an Artemis System (Eigen Health, Grass Valley, California) and a Hitachi Ultrasound scanner utilized during the MRI-TRUS fusion biopsy performed for each patient. Conventional brightness-modulated (b-mode) TRUS images of the prostate were acquired using 2D end-fire probes with center frequencies of 7.5-10 MHz and reconstructed to form 3D prostate image volumes. The MRI-US fusion biopsy involved targeting the suspicious lesions outlined by radiologists on MRI sequences and projected to the TRUS images by the Artemis system. The biopsy procedure further continued with the systematic biopsy of 12-14 locations for each patient.

**Ground Truth:** For the biopsy cohorts $\boldsymbol{C}_{BX}^{Stanford}$ and $\boldsymbol{C}_{BX}^{UCLA}$, the cancer ground truth was obtained from targeted lesions outlined by radiologists that were then confirmed through pathology lab testing of their biopsy samples. We assigned each lesion as cancer ground truth based on the highest International Society of Urological Pathology[35] (ISUP) grade group (GG) found within the lesion. The RP cohort $\boldsymbol{C}_{RP}^{Stanford}$ included patients from whom we obtained whole-mount pathology slides with pixel-level indication of the cancer presence. Following pathology results, an expert image analysis scientist (MR, > 15 years of experience with registering prostate pathology and radiology) manually outlined the extent of cancer on TRUS images of each patient using the whole mount pathology reference slides.

**Data Preprocessing:** MRI image sequences were acquired using different scanners and protocols; therefore, we preprocessed them to a similar spacing, size, and intensity range. The T2w, ADC, and DWI sequences were resampled to 0.5mm x 0.5mm x 3.0mm/voxel using B-Spline interpolation. Moreover, we center-cropped the x-y plane of the MRI sequences to a size of 128mm x 128mm, corresponding to an image size of 256 x 256 pixels, and padded the image volumes if needed. We performed a similar preprocessing on TRUS image volumes, except that we resampled the TRUS voxels to a spacing of 0.5mm x 0.5mm x 0.5mm/voxel due to the higher resolution TRUS frames in the coronal plane. Additionally, we normalized all the intensity sequences using Z-score normalization in refence to the prostate gland to ensure that the pixel intensities within each gland have a mean of zero and standard deviation of one for each sequence. To enable our model to eventually learn to natively identify prostate cancer in the TRUS space, we utilized the deep learning model previously detailed in[36] to perform 3D affine registration and register the MRI sequences to the TRUS space before using them as input to our AI model.

### *4.2. AI Model Description*

Our AI model is based on the well-known nnUNet framework[37], which uses a UNet architecture to extract multi-scale features at various resolutions and relies on expanding and contracting paths to distill knowledge across each scale[38,39]. We trained the model using 3D image volumes where each volume was automatically divided into 3D voxel patches for training. We designed the model architecture to treat each image sequence as a separate input channel, from which 3D voxel patches were extracted and concatenated. These concatenated patches were processed through a unified encoder, which progressively reduced the resolution to extract hierarchical latent representations at multiple scales. The lowest-scale latent representations were then passed to a unified decoder, which progressively increased the resolution to reconstruct feature representations associated with discriminatory characteristics that ultimately help identify cancerous pixels. At each decoding stage, encoded latent representations from the corresponding encoder stage were incorporated via skip connections, which merge spatially aligned feature maps from the encoder and decoder to enhance feature preservation.

Each encoder and decoder stage consists of two 3D convolutional layers with a kernel size of 3x3x3, followed by 3D instance normalization and a Leaky ReLU (rectified linear unit) activation. Encoding stages are further followed by a downsampling operation with strides of 2x2x2, except in the first stage where no downsampling is applied. Decoding stages include the concatenation of latent representation from skip connections, similar convolutional and normalization/activation layers as in the encoder, and an upsampling operation with strides of 2x2x2 to restore spatial resolution. Finally, segmentation heads consisting a 3D convolutional layer with a kernel size of 1x1x1 was applied to produce class scores for each voxel, followed by a SoftMax activation to assign probabilities to any or all of the following labels: 1) prostate gland, 2) any cancer (GG $\geq$ 1), or 3) CsPCa (GG $\geq$ 2). We empirically found that enabling the model to learn all the above labels was helpful in achieving a higher performance in identifying CsPCa. Binary cross entropy loss and dice loss were combined as the loss objective to train the model. We used the same 3D UNet backbone for the baseline unimodal MRI and TRUS models, where the number of input channels was based on the number of image sequences provided by each modality. An overview of the model architecture is illustrated in Figure 1B.

### *4.3. Experimental Design*

To better measure the generalizability of our multimodal AI model, we only utilize 1410 cases from the $\boldsymbol{C}_{BX}^{Stanford}$ cohort for training and validation during model development. We utilize the remaining 345 cases from this cohort along with all the cases from $\boldsymbol{C}_{BX}^{UCLA}$ and $\boldsymbol{C}_{RP}^{Stanford}$ cohorts as independent test cohorts for a total of 1700 test cases. The details of each cohort, including the distribution of patients without cancer vs. patients with indolent cancer or CsPCa within each cohort are shown in Figure 1A.

**Multimodal Model vs. Unimodal Models:** To examine the role of each imaging modality in enabling the AI model to learn to identify prostate cancer, we train our 3D UNet model using three different setups, including 1) a unimodal model utilizing TRUS images as input, 2) a unimodal model utilizing MRI sequences as separate input channels, and 3) a multimodal model utilizing both TRUS and MRI sequences as separate input channels. In this experiment, we evaluate the performance of each model in identifying CsPCa in the TRUS space against the pathology-confirmed lesion ground truths from each test cohort. The training of the unimodal MRI model was done in the MR image space which empirically yielded the best performance, then the prediction probabilities for each test case was projected to the TRUS space using the same registration transformation obtained during the preprocessing of each MRI case.

**Multimodal Model vs. Radiologists:** We further evaluate the performance of our multimodal model as well as the performance of radiologist readings during multidisciplinary routine clinical care against the cancer ground truths obtained from whole-mount pathology slides captured from radical prostatectomy patients in the $\boldsymbol{C}_{RP}^{Stanford}$test cohort. Since whole mount pathology slides provide cancer annotations for each pixel within the image, this head-to-head evaluation is extremely reliable and perhaps the closest measure to the real-world performance of the AI model. Since the radiologist readings are originally performed on MR images, we utilize the same registration transformation obtained during the preprocessing of each MRI case to project the radiologist lesion outlines to the TRUS space for evaluation against the cancer ground truths in that space and enable a case-by-case performance evaluation against the multimodal model.

### *4.4. Performance Evaluation*

To assess the performance of each AI model against the ground truth, we performed a lesion-level evaluation[18,40]. While the models were trained to learn the three class labels, we focus the evaluation on assessing the ability of the models in identifying CsPCa lesions due to its clinical relevance, as well as considering negatives at a sextant level to assess the model prediction specificity. We report area under the Receiver Operating Characteristics curve (ROC), area under the Precision Recall curve (PR), sensitivity, specificity, and negative predictive value (NPV). Additionally, we compute the overlap between the cancer ground truth and predicted lesions using 1) Overall Dice that captures the overlap between all predictions and the ground truth, 2) Lesion Dice which captures the overall dice for patients with correctly predicted cancers.